\documentstyle[aps,floats,epsfig]{revtex}

\def\deg{\ifmmode{^{\circ}}\else ${^{\circ}}$\fi}
\def\bi{\begin{itemize}}
\def\ei{\end{itemize}}
\def\ed{\end{document}}

\def\cf#1{\ifmmode{\cal #1}\else${\cal #1}$\fi}

\def\be{\begin{equation}}
\def\ee{\end{equation}}
\def\beas{\begin{eqnarray*}}
\def\eea{\end{eqnarray}}
\def\bea{\begin{eqnarray}}
\def\eeas{\end{eqnarray*}}
\def\tfrac#1#2{{\textstyle\frac{#1}{#2}}}

\def\gev{\ifmmode{\mbox{GeV}}\else GeV\fi}

\def\ton{T_{\rm on}}

\begin{document}

\twocolumn
\renewcommand{\topfraction}{1.0}
\twocolumn[\hsize\textwidth\columnwidth\hsize\csname
@twocolumnfalse\endcsname
\title{Extragalactic Sources for Ultra High Energy Cosmic Ray Nuclei}
\author{Luis Anchordoqui, Haim Goldberg, Stephen Reucroft, and John Swain}
\address{Department of Physics, Northeastern University, Boston, MA 02115, USA}

\maketitle

\begin{abstract}

In this article we examine the hypothesis that the highest energy cosmic rays
are complex nuclei from extragalactic sources.
Under reasonable physical assumptions, we show that the nearby metally
rich starburst galaxies (M82 and NGC 253) can produce all the events
observed above the ankle. This requires diffusion of particles below
$10^{20}$~eV in extragalactic magnetic fields $B \approx 15$~nG.
Above $10^{19}$~eV, the model
predicts the presence of significant fluxes of medium mass and heavy nuclei
with small rate of change of composition.
Notwithstanding, the most salient feature of the starburst-hypothesis is
a slight anisotropy induced by iron debris just before the spectrum-cutoff.

\end{abstract}

\vskip2pc]

\section{Introduction}

The cosmic ray (CR) spectrum above $10^{10}$ eV (where the Sun's magnetic
field is no longer a concern) can be described by a series of power
laws with the flux falling about 3 orders of magnitude for each decade
increase in energy  \cite{reviews}.
Above $10^{14}$ eV, the flux becomes so low that
direct measurements using sophisticated equipment on satellites or high
altitude balloons are limited in detector area and in exposure
time. Ground-based experiments with large apertures make such a low
flux observable after a magnification effect in the upper
atmosphere: the incident cosmic radiation interacts with
atomic nuclei of the air molecules and produces extensive air
showers which spread out over large areas. Continuously running monitoring
through ingenious installations has raised the maximum observed primary
particle's energy to higher than $10^{20}$~eV \cite{snowmass}.

While theoretical subtleties surrounding CR acceleration provide ample
material for discussion, the debate about the origin of CRs up to the knee
($\sim 10^{15.5}$ eV) has reached a consensus that they are produced in
supernova explosions \cite{supernova}. The change of the spectral index
(from $-2.7$ to $-3.0$) near the knee, presumably reflects a change in
origin and the takeover of another, yet unclear type of source. The
spectrum steepens further to $-3.3$ above $\sim 10^{17.7}$~eV (the dip)
and then flattens to an index of $-2.7$ at $\sim 10^{18.5}$~eV (the ankle).
A very widely held interpretation of the modulation features is that above
the ankle
a new population of CRs with extragalactic origin begins to dominate the more
steeply falling Galactic population \cite{AGASA}. The origin of the
extragalactic channel is somewhat mysterious.

CRs do not travel unhindered through intergalactic space, as
there are several processes that can degrade the particles' energy. In
particular, the thermal photon background becomes highly blue shifted
for ultrarelativistic protons. The reaction sequence
$p\gamma \rightarrow \Delta^+ \rightarrow \pi^0 p$ effectively degrades the
primary proton energy providing a strong constraint on the proximity of
CR-sources, a phenomenon known  as the
Greisen-Zatsepin-Kuz'min (GZK) cutoff \cite{gzk}. Specifically, fewer
than 20\% of
$10^{20.5}$~eV ($10^{20}$~eV) protons can survive a trip of 18~Mpc
(60~Mpc) \cite{new_gzk}.
A heavy nucleus undergoes photodisintegration in the microwave and infra-red
backgrounds; as a result, iron nuclei do not survive fragmentation over comparable
distances \cite{nucleus}. Ultra high energy gamma rays would travel even
shorter paths due to pair production on radio photons \cite{gamma}.

In order to analyze the effect of energy losses in the observed
spectrum, it is convenient to introduce the accumulation factor
$f_{\rm acc}$, defined as the ratio of energy-weighted fluxes for
low ($10^{18.7}$ eV -- $10^{19.5}$ eV) and high ($> 10^{20}$ eV)
energy CRs above the ankle. In the case where the cosmic rays are
protons from a uniform distribution of sources active over
cosmological times, the cutoff due to the photopion processes
relates the accumulation factor to a ratio of the GZK distances
\cite{FP} and leads to $f_{\rm acc} \sim 100$. A similar value
for $f_{acc}$ is obtained for nuclei due to photodisintegration.
Therefore, in the case of ordinary baryonic CRs, if Earth is
located in a typical environment and all CR-sources have smooth
emission spectra, the observed spectrum above the ankle should
have an offset in normalization between low and high energy given
by $f_{\rm acc}$. To reproduce the recorded spectrum (i.e.,
$f_{\rm acc} \sim 1$), the power of nearby sources (say, $10$ Mpc
or so) should be comparable to that of all other sources
(redshift $z>0.5$) added together. This condition imposes stern
constraints on models describing the origin of baryonic ultra
high energy CRs. For instance, ``top down'' models (with hard
injection spectra $\propto E^{-1}$) \cite{top_down} would fail to
reproduce the detected population of CRs below the GZK energy by
more than an order of magnitude. For models that rely  on
GZK-evading messengers \cite{gzk_evading}, $f_{\rm acc}$ depends
on the details of the model. In the case of messengers which can
induce showers across the entire energy spectrum, one expects
enhancement on the low energy side only from the baryonic
component,  and $f_{\rm acc}$ depends on the interaction length in
the cosmic microwave background and on the relative energy spectra
at the source. For messengers whose attenuation length is
comparable to the horizon, and which do not shower at the lower
energies, $f_{\rm acc}\ll 1.$  Summing up, the smoothness of the
observed CR spectrum suggests, as the simplest explanation, that
nearby sources should be significantly more concentrated or more
powerful than average.

On a different track, any candidate model addressing the origin of ultra high
energy CRs should properly match the main features of the observed
extensive air showers. As the cascade develops in the atmosphere, the
number of particles in the shower increases until the
secondary particles' energy is degraded to the point where ionization
losses dominate, and the density of particles starts to decline.
The number of particles as a function of the amount of atmosphere
penetrated by the cascade in g~cm$^{-2}$ becomes a smooth curve, the
so-called ``longitudinal profile''. The atmospheric depth at which
the shower reaches its maximum size is referred to as the depth of shower
maximum $X_{\rm max}$, and is often regarded as the most basic parameter
of the shower. It increases with primary energy as more cascade generations
are required in the cooling of secondary products. For a given total energy
$X_{\rm max}$ is related to the energy per nucleon of the shower progenitor.
Unfortunately, extracting information on the nature of the primaries from
the air shower they produce has proved to be exceedingly difficult.
The most fundamental drawback is that the first few cascade steps
are subject to large inherent fluctuations and consequently this limits the
event-by-event mass resolution of the experiments. In addition,
the center of mass energy of the first interactions is well beyond those
reached in collider experiments. Thus, one needs to rely on hadronic
interaction models that attempt to extrapolate (using different mixtures of
theory and phenomenology) our understanding of particle physics.

\begin{figure}
\label{s1}
\begin{center}
\epsfig{file=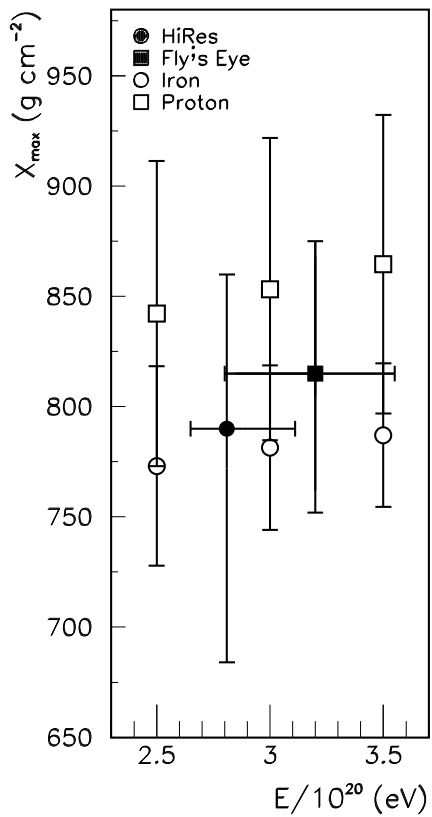,width=6.cm,clip=}
\caption{Simulated depth of shower maxima of two different particle species
superimpossed over experimental data.}
\end{center}
\end{figure}

An analysis of the histogram of $X_{\rm max}$, observed by the
Fly's Eye experiment, indicates that there is a significant fraction of nuclei
with charge greater than unity in the energy range ($10^{18.5}$ eV --
$10^{19}$ eV, and somewhat above) \cite{ww}. To examine the situation above
the GZK-energy, simulations of giant air shower evolution have been
performed by means of the code {\sc airesq} (version 2.1.1) \cite{sergio}.
Several sets of protons and iron nuclei were injected at 100 km above sea
level. The geomagnetic field was set to reproduce that in the Utah desert. All
shower particles with energies above the following
thresholds were tracked: 750 keV for gammas, 900 keV for electrons and
positrons, 10 MeV for muons, 60 MeV for mesons and 120 MeV for nucleons
and nuclei. The charged multiplicity, essentially electrons and positrons,
was used to determine the number of charged particles and the location of the
shower maximum by means of 4-parameter fits to the Gaisser-Hillas
function. In Fig. 1 we show the evolution of
$X_{\rm max}$ above the GZK-energy for protons and iron nuclei.
For comparison, we also show the depth of shower maxima of the two highest
energy events recorded by Fly's Eye and HiRes experiments \cite{utah}.
The observed values of $X_{\rm max}$ are consistent with both proton and
iron primaries, perhaps (speculatively) suggestive of a medium mass nucleus
\cite{halzen}.

In the search for the trans-GZK-sources another observable that one
has to take into account is the CR arrival directions. The observed events
above $\approx 10^{20}$ eV are distributed widely over the sky, with no
plausible counterparts (such as sources in the Galactic Plane or in the
Local Supercluster). Moreover, the data are consistent with an isotropic
distribution of sources in sharp contrast  to the anisotropic distribution
of light within 50 Mpc. At first glance, this seems to contradict any
explanation based on nearby sources. However, if the highest energy CRs are
heavy nuclei,
one cannot yet rule out that extragalactic/galactic magnetic fields could
tangle up the particle paths, camouflaging the exact location of the sources.
Intergalactic field strengths and coherent lengths are not well established,
but it is plausible to assume that fields have coherent directions on scales
of $\ell \sim 0.5 - 1$~Mpc. A recent estimate on the extragalactic
magnetic field
in the neighborhood of the Milky Way suggests $B> 10$ nG \cite{azucar}.
However, magnetic fields of a few $\mu{\rm G}$ structured in cells of
$\sim 1$ Mpc  cannot be excluded \cite{ckb}. Such a large field would
completely deflect trans-GZK proton orbits. For a CR nucleus
of charge $Ze$ in a
magnetic field $B_{\rm nG} = B/10^{-9}$~G, the Larmor radius is
\begin{equation}
R_{\rm L} \approx \frac{E_{18}}{Z \,B_{\rm nG}}\,\,{\rm Mpc}\,\,,
\end{equation}
where $E_{18} = E/10^{18}$ eV. Therefore, the assumption that the
giant air showers $E > 10^{20}$ eV were triggered by
heavy nuclei implies  ordered ($\ell \sim 1$ Mpc)
extragalactic magnetic fields $B_{\rm nG} < 15$ (at least in the
outskirts of the Galaxy), or else nuclei would be
trapped in magnetic subdomains suffering catastrophic
spallations \cite{beo_beo}.

In a previous paper \cite{cen_a}, there was explored the
hypothesis that CRs above the ankle are (mostly) protons from the
nearby radio galaxy Centaurus A \cite{FP}, accelerated in a ``hot
spot'' in the northern middle lobe. These protons were assumed to
diffuse in a magnetic field of ${\cal O}(\mu$G) over the transit
distance $\sim 3.4$ Mpc to Earth. In this work, in view of the
possibility raised by the air shower profiles, we examine an
alternate hypothesis: that the composition of CRs above the ankle
are largely heavy nuclei which originate in two nearby sources,
and traverse extragalactic magnetic fields which conform to the
restriction outlined in the previous paragraph. As will be seen,
this hypothesis (as well as the previous one involving Cen A),
will be subject to specific testing in the coming array of high
statistics cosmic ray observations.

\section{Starburst-hypothesis}

If the trans-GZK particles are heavy nuclei, then the nearby
($\sim 3$ Mpc \cite{heckman}) starburst galaxies M82 ($l=141^\circ,
b=41^\circ$) and NGC 253 ($l=89^\circ, b= -88^\circ$) would probably
be the sources of most ultra high energy CRs observed
on Earth. Starbursts
are galaxies undergoing a massive and large-scale star formation episode.
Their characteristic signatures are strong infrared emission (originated in
the high levels of
interstellar extinction), a very strong HII-region-type emission-line
spectrum (due to a large number of O and B-type stars), and a considerable
radio emission produced by recent supernova remnants (SNRs). Typically,
the starburst region is confined to the
central few hundreds of parsecs of the galaxy, a region that can be
easily 10 or more times brighter than the center of normal spiral galaxies.
In the light of such a concentrated activity, the
existence of  galactic superwinds is not surprising\cite{heckman}.

Galactic-scale superwinds
are driven by the collective effect of supernovae and massive star winds. The
high supernovae rate creates a cavity of hot gas ($\sim10^8$ K) whose
cooling time is much greater than the expansion time scale. Since the wind
is sufficiently powerful, it can blow out the interstellar medium of the
galaxy avoiding it remaining trapped as a hot bubble. As the cavity
expands a strong shock front is formed on the contact surface with the
cool interstellar medium. The shock velocity can reach several thousands
of kilometers per second and ions like iron nuclei can be then efficiently
accelerated in this scenario up to ultra high energies
by Fermi's mechanism \cite{ngc}.

In a first stage, ions are diffusively accelerated  at
single supernova shock waves within the nuclear region of the galaxy.
Energies up to $\sim 10^{15}$ eV can be achieved in this step \cite{supernova}.
Heavy nuclei are not photodissociated in the process despite
the large photon energy densities (mostly in the far infrared) measured
in the central region of the starburst. The escape of the CR outflow is
convection dominated. In fact, the
presence of several tens of young SNRs with very high expansion velocities
and thousands of massive O stars
(with stellar winds of terminal velocities up to 3000 km s$^{-1}$) must
generate collective plasma motions of several thousands of km per second.
Then, due to the coupling of the magnetic field to the hot plasma, the
magnetic field is also lifted outwards and forces the CR gas to
stream along from the starburst region. Most of the nuclei escape
in opposite directions along the symmetry axis of the system, as the total
path traveled is substantially shorter than the mean free path \cite{ngc}.

Once the nuclei escape from the central region of the galaxy (with energies
of $\sim 10^{15}$ eV) they are injected into the galactic-scale wind and
experience further acceleration at its terminal shock.
For this second step in the acceleration process, the photon field energy
density drops to values of the order of the microwave background radiation
(we are now far from the starburst region), and consequently, iron nuclei
are safe from photodissociation while energy increases from $\sim
10^{15}$ to $10^{20}$ eV. In terms of parameters that can be determined
from observations, the nucleus maximum energy is given by \cite{ngc},
\begin{equation}
E_{\rm max}\approx\frac{1}{2}\, Ze\,B \,
\frac{\dot{E}_{\rm sw}}{\dot{M}}\, \ton\,\,,
\label{14}
\end{equation}
where  $\dot{E}_{\rm sw} \sim 2.7 \times 10^{42}$ erg s$^{-1}$ is the
superwind kinetic energy flux and  $\dot{M} = 1.2
{\rm M}_{\odot}$ yr$^{-1}$ is the mass flux generated by the
starburst \cite{heckman}.
The  age $\ton$ can be estimated from numerical models that use theoretical
evolutionary tracks for individual stars and make sums over the entire stellar
population at each time in order to produce the galaxy luminosity as a
function of time \cite{edades}. Fitting the observational data, these
models provide a range of suitable ages for the starburst phase that
goes from $50$~Myr to $160$ Myr \cite{edades}. These models
must assume a given initial
mass function (IMF), which usually is taken to be a power-law with a variety of
slopes. Recent studies have shown that the same IMF can account for the
properties of both NGC 253 and M82 \cite{engelbracht}.
Besides, a region (referred to as M82 ``B'') near the galactic center of
M82, has been under suspicion  to be a fossil starburst site in which an
intense episode of star formation occurred over 100~Myr ago \cite{fossil}.
The derived age distribution suggests steady, continuing cluster formation at
a modest rate at early times ($> 2$ Gyr ago), followed by a concentrated
formation episode 600~Myr ago and more recent suppression of cluster
formation. In order to get some
estimates on the maximum energy, let us assume $B\sim50\mu$G,
a choice consistent with observation \cite{paglione}.
Replacing in Eq.~(\ref{14}) in favor of these figures, already for
$\ton = 50$ Myr one obtains
\begin{equation}
E_{\rm max}^{\rm Fe} > 10^{20}\;\;\;\;{\rm eV}.
\end{equation}

Now, we could use the rates at which starbursts inject mass, metals and
energy into superwinds to get an estimate on the CR-injection spectra.
Let us introduce $\epsilon$, the efficiency of ultra high energy CR
production by the superwind kinetic energy flux.
Using equal power per decade over the interval $10^{18.5}\,{\rm eV}<
E< 10^{20.6}\,{\rm eV}$, we obtain a source CR-luminosity
\begin{equation}
\frac{E^2 \,dN_0}{dE\,dt} \, \approx 3.5 \,\epsilon \,10^{53}
{\rm eV/s} \label{1}
\end{equation}
where the subscript ``0'' refers to quantities at the source. The
density of CRs at the present
time $t$ of energy $ < 10^{20}$~eV at a distance $r$ from a source
(assumed to be continuously emitting at a constant spectral
rate $dN_0/dE\,dt$ from time $t_{\rm on}$ until the present)
is \cite{cen_a}
\begin{eqnarray}
\frac{dn(r,t)}{dE} & = & \frac{dN_0}{dE\,dt}
\frac{1}{[4\pi D(E)]^{3/2}} \int_{t_{\rm on}}^t dt'\,
\frac{e^{-r^2/4D(t-t')}}{(t-t')^{3/2}}  \nonumber \\
 & = & \frac{dN_0}{dE\,dt}
\frac{1}{4\pi D(E)r} \,\,I(x),
\label{2}
\end{eqnarray}
where $D(E)$ stands for the diffusion coefficient,
$x = 4D\ton/r^2 \equiv \ton/\tau_D$, $\ton=t-t_{\rm on},$ and
\begin{equation}
I(x) = \frac{1}{\sqrt{\pi}} \int_{1/x}^\infty \frac{du}{\sqrt{u}} \,\,
e^{-u}\ \ .
\end{equation}
In each ``scatter'', the diffusion coefficient describes an
independent angular deviation of particle trajectories whose
magnitude depends on the Larmor radius. CRs with
energies $E < 10^{18}\, \ell_{\rm Mpc}\, Z B_{{\rm nG}}$ eV remain
trapped inside cells of size $\ell_{\rm Mpc}= \ell/(1\ {\rm Mpc})$,
attaining efficient
diffusion when the wave number of the associated Alfv\'en wave is equal to
the gyroradius of the particle \cite{wentzel-drury}.  It
may be plausible to assume a Kolmogorov form for the turbulent
magnetic field power spectrum, this gives for a diffusion
coefficient\cite{blasi-olinto}
\begin{equation}
D(E) \approx 0.048 \left( \frac{E_{18} \,\ell^2_{{\rm
Mpc}}}{Z\,B_{{\rm nG}}} \right)^{1/3} \,{\rm Mpc}^2/{\rm Myr}.
\end{equation}
For
$\ton\rightarrow \infty$, the density approaches its time-independent
equilibrium value $n_{\rm eq}$, while for $\ton= \tau_D=r^2/4D$,
$n/n_{\rm eq} = 0.16$.
To further constrain the parameters of the model, we  evaluate the
energy-weighted approximately isotropic proton flux at  $10^{19}$ eV,
which lies in the center of the flat
``low energy'' region of the spectrum:
\begin{eqnarray}
E^3 J(E) & =
& \frac{Ec}{(4\pi)^2d\,D(E)} \frac{E^2 \,dN_0}{dE\,dt}\,
I_\star  \nonumber \\
 & \approx & 2.3 \times 10^{26} \, \epsilon \,
I_\star
\, {\rm eV}^2 \, {\rm
m}^{-2} \, {\rm s}^{-1} \, {\rm sr}^{-1}, \label{jp}
\end{eqnarray}
where $I_\star = I_{\rm M82} + I_{\rm NGC\ 253}$.
In the second line of the equation we used
$B_{\rm nG} = 15$, $\ell_{\rm Mpc} = 0.5$, and an average $Z=20$.
We fix
\begin{equation}
\epsilon\, I_\star =0.013,
\label{mimi}
\end{equation}
after comparing Eq.(\ref{jp}) to the observed CR-flux: $E^3
J_{\rm obs}(E)  = 10^{24.5}$ eV$^2$ m$^{-2}$ s$^{-1}$ sr$^{-1}$
\cite{reviews}. Note that the contribution of $I_{\rm M82}$ and
$I_{\rm NGC\ 253}$ to $I_\star$ critically depends on the age of
the starburst. In Fig. 2 we show the relation
``starburst-age/superwind-efficiency'' derived from Eq.
(\ref{mimi}). We have assumed that both M82 and NGC 253 were
active for $115$ Myr ($\epsilon \approx 10\%$), beyond this epoch
CR-emission must be associated to M82 ``B''.

\begin{figure}
\label{s2}
\begin{center}
\epsfig{file=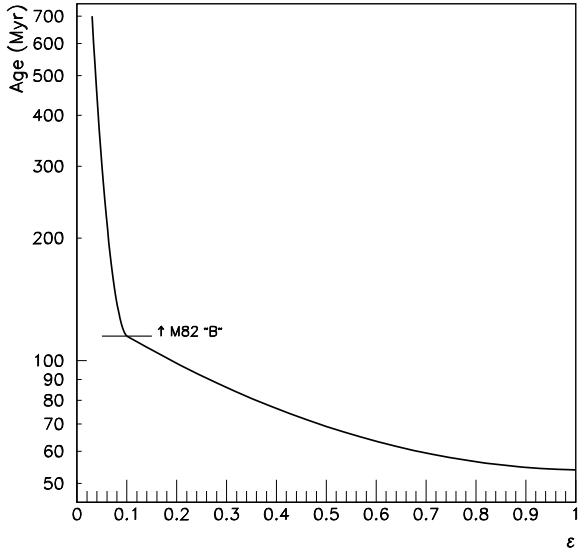,width=8.5cm,clip=}
\caption{Age of the starbursts as a function of the efficiency
of CR-production $\epsilon$.}
\end{center}
\end{figure}

Above $> 10^{20.2}$ eV iron nuclei do not propagate diffusively.
Moreover, the CR-energies get attenuated
by photodisintegration off the microwave background radiation and the
intergalactic infrared background photons. The disintegration rate of
$^{56}$Fe as a function of the Lorentz factor $\Gamma$ can
be parametrized as follows
 \cite{prd2},
\begin{mathletters}
\begin{eqnarray}
R(\Gamma) & = & 3.25 \times 10^{-6}\,
\Gamma^{-0.643} \nonumber \\
 & \times &
\exp (-2.15 \times 10^{10}/\Gamma)\,\, {\rm s}^{-1}\,\,,
\label{par1}
\end{eqnarray}
if $\Gamma \,\in \, [1. \times 10^{9}, 3.68 \times 10^{10}]$, and
\begin{equation}
R(\Gamma) =1.59 \times 10^{-12} \,
\Gamma^{-0.0698}\,\, {\rm s}^{-1} \,\,,
\label{par2}
\end{equation}
if $ \Gamma\,
\in\,
[3.68 \times 10^{10}, 1. \times 10^{11}]$.
\end{mathletters}
At this stage it is worthwhile to point out that the knowledge of
$^{56}$Fe effective nucleon loss rate is enough to obtain the
corresponding value of $R$ for any other nuclei \cite{nucleus},
\begin{equation}
\left. \frac{dA}{dt}\right|_A \sim \left. \frac{dA}{dt}\right|_{\rm Fe}
\left(\frac{A}{56}\right).
\label{avga}
\end{equation}
This means that $A(t)$ is an exponential function of time with an
$e$-folding time of $56\, \left. dA/dt\right|_{\rm Fe}^{-1}$.
Now, since the emission of nucleons is isotropic
in the rest frame of the nucleus, the average fractional energy loss
is equal to the fractional loss in mass number of the nucleus (i.e.,
$\Gamma$ is conserved). The relation that determines the energy attenuation
length of iron as a function of the time flight $t$ is then
\begin{eqnarray}
E (t) & = & 938\,\, A(t)\,\, \Gamma\,\,\, {\rm MeV}  \nonumber \\
  & = & E_0\, \exp\left[\frac{-R(\Gamma)\,t}{56}\right],
\label{phd_f}
\end{eqnarray}
where $E_0 \equiv 938\,\, A_0\,\, \Gamma$ MeV, denotes the
nucleus'
emission energy \cite{phd}.  This relation
imposes a {\it strong} constraint on the location of
nucleus-sources: note that {\it less than 1\%
of iron nuclei (or any surviving fragment of their spallations)
can survive more than $3 \times 10^{14}$ s with an energy $>10^{20.5}$ eV}.

In the non-diffusive regime, the accumulated deflection angle $\theta(E)$
from the direction of the source, located at a distance
$d$, can be estimated
assuming that the particles make a random walk in the magnetic field \cite{we}
\begin{equation}
\theta(E) \approx 0.54^\circ \,\left (\frac{d}{1\,{\rm Mpc}}\right)^{1/2}\,
\left(\frac{\ell}{1\,{\rm Mpc}}\right)^{1/2}\,
\frac{Z\, B_{\rm nG}}{E_{20}},
\end{equation}
where $E_{20} \equiv E/10^{20}$ eV. Therefore, if $B \sim 15$~nG all
directionality is lost. The resulting time delay with respect to
linear propagation is given by
\begin{equation}
\tau_{\rm delay}(E) \approx \frac{d\,\, \theta^2}{4\,c}.
\end{equation}
and the total travel time is
\begin{equation}
t\approx \frac{d}{c}\ \left(1+\tfrac{1}{4}\theta^2\right)\ \ .
\label{time}
\end{equation}

As an example, we apply these consideration to the highest energy
Fly's Eye event. Including statistical and systematic
uncertainties, the energy of this event is $3.2\pm 0.9\times
10^{20}$ eV. Eqs. (\ref{par1}) and (\ref{phd_f}) relate the
uncertainty in energy to the uncertainty in the attenuation time:
\begin{equation}
\frac{\delta t}{t}\simeq \left(\frac{11.3}{E_{20}}\right)\,\,
\left(\frac{\delta E}{E}\right)\ \ .
\end{equation}
>From these considerations, we find that the upper limit on the transit time for
a nuclear candidate for the highest energy Fly's Eye event is $\sim 6\times 10^{14}$ s.
The arrival
direction of the highest energy
Fly's Eye event is  $37^{\circ}$ from M82 \cite{elbert}. With
$d\simeq 3$ Mpc and $\theta=37^{\circ},$ we find
from Eq. (\ref{time}) a transit time $t\simeq 3.4\times 10^{14}$ s, well
within the stated upper limit \cite{yakutsk}.

For average deflections of $60^\circ$, the time of flight is
$\sim 3.9 \times 10^{14}$ s, and consequently there is a sharp end of the
CR-spectrum near the maximum observed energy. Indeed, this leads to a
slight anisotropy just before the cutoff, and eventually to a north-south
asymmetry on the tail \cite{asymmetry}.
It is rather difficult to
assess whether events with energies $>10^{20.5}$ eV are plausible. This would
require an event-by-event analysis because the maximum energy
strongly depends on $\tau_{\rm delay}$. We do not attempt to make yet another
estimate in the present article and only note that the energy-weighted flux
beyond the GZK-energy (due to a single M82 flare) \cite{prd1}
\begin{eqnarray}
E^3 J(E) & =
& \frac{E}{(4 \pi d)^2} \frac{E_0^2 \,dN_0}{dE_0\,dt}\,e^{-R\,t/56}
\nonumber \\ & \approx & 2.7 \times 10^{25} E_{20}
\epsilon \,e^{-R\,t/56}\,
{\rm eV}^2 \, {\rm m}^{-2} \, {\rm s}^{-1} \, {\rm sr}^{-1},
\end{eqnarray}
is easily consistent with observation \cite{reviews}. The analytical
study outlined in this paper should be followed up by numerical studies,
especially to refine our estimates above the GZK-energy.

\section{Conclusion}

We have shown that the nucleus-emitting-sources of our
backyard have enough power to produce all CRs observed above the ankle.
Starburst galaxies can accelerate iron nuclei above the GZK-energy if
a two-step process is
involved. The crucial point is that for energies $>10^{15}$ eV,
acceleration occurs in the terminal shock of the starburst superwind,
well outside the problematic central region.

Below $10^{19}$ eV, the distribution of the CR arrival directions
is expected to be completely isotropic because of Kolmogorov
diffusion in ordered ($\ell_{\rm Mpc} =0.5$) extragalactic
magnetic fields $10 <B_{\rm nG}< 15$. In addition,
(de)magnification of the fluxes by lensing effects are expected
due to deflections in the regular Galactic magnetic field
\cite{toes}. Furthermore, medium mass and heavy nuclei with
energies $<10^{19.7}$ eV would also have an isotropic
distribution in the sky.

On the other hand, ultra high energy ($E>10^{19.7}$~eV) light
nuclei ($Z <10$) do not propagate diffusively. However, as can be seen in
Eq.(\ref{14}), light nuclei are not
copiously accelerated to ultra high energies in the starburst. Note also that
the nucleons ($\alpha$-particles) emitted in the photodisintegration
process have energies well below $10^{19}$~eV and consequently do not
produce any anisotropy.

It should be noted that several groups \cite{clusters} have recently
reported evidence of clustering (6 doublets and 1 triplet, with the chance
probability $7 \times 10^{-4}$ \cite{agasa_clusters}). Magnetic
focussing in the magnetic field structure could -- in principle -- account
for directional clustering to explain the current sparse data \cite{lsb}.
However,
if not a statistical fluctuation, and clusters are well established in very
much larger data sets, they would constitute a serious objection to the
model.

All in all, within this scenario almost all CRs above the
GZK-energy would be medium mass and heavy nuclei, yielding a
small rate of change of composition (in agreement with
observation \cite{agasa}). In addition, the model predicts a
slight anisotropy above $10^{20.4}$~eV produced by non-diffusive
iron debris. This makes the model capable of being proven false.
The limited statistics in the observed data make it impossible to
definitively test the ``starburst-hypothesis'' at this time. The
coming avalanche of high quality CR-observations \cite{reviews}
promisses to give the final verdict on these speculations.

\hfill

\acknowledgments{
The work was partially supported by CONICET (Argentina) and the
National Science Foundation (USA).
}

\end{document}